\RequirePackage{fix-cm}
\documentclass[smallextended]{svjour3}       
\smartqed  
\usepackage{graphicx}
\usepackage[numbers]{natbib}
\setcitestyle{square}
\begin{document}

\title{Is there a Cosmological Basis for $E=mc^2$?
}
\subtitle{}

\titlerunning{Is there a Cosmological Basis for $E=mc^2$?}        

\author{Geraint F. Lewis
}


\institute{Geraint F. Lewis \at
              Sydney Institute for Astronomy, \\
              School of Physics, A28, \\
              The University of Sydney, \\
              NSW 2006, Australia \\
              \email{geraint.lewis@sydney.edu.au}           %
}

\date{Received: date / Accepted: date}

\maketitle

\begin{abstract}
It has recently been claimed that relativity's most famous equation, $E=mc^2$, has a cosmological basis, representing the gravitational binding energy for a particle to escape from the origin to a gravitational horizon of the universe. 
In this paper, I examine these claims in detail, concluding that they result from a misinterpretation of motion of particles in the cosmological space-time, and an incorrect application of 4-vectors.  
Finally, I demonstrate that the origin of $E=mc^2$ comes from its usual relativistic interpretation, namely that it is the energy of a particle as seen in its own rest-frame. 
\end{abstract}

\section{INTRODUCTION }
\label{sec:intro}

Recently, it  has claimed that one of fundamental relationships from relativity, namely $E=mc^2$, has a cosmological basis, interpreting it as ``gravitational binding energy'', the energy needed for a particle to escape from the origin to a gravitational horizon \cite{2019IJMPA..3450055M}. 
In this paper, I briefly outline that these conclusions are erroneous, resulting from a misinterpretation of the form of the metric and the normalisation of 4-vectors in general relativity.

The layout of this paper is as follows; Section~\ref{sec:origin} presents the arguments laid out in support of a cosmological origin of $E=mc^2$ \cite{2019IJMPA..3450055M}, whilst Section~\ref{sec:nobasis} demonstrates the erroneous nature of some of these clams. Section~\ref{sec:conclusions} presents the brief conclusions to this work.

\section{A Cosmic Basis for $E=mc^2$?}
\label{sec:origin}

The starting point is the Friedmann-Lemaitre-Robertson-Walker (FLRW) metric for a spatially flat universe, with an interval given by;
\begin{equation}
ds^2 = c^2 dt^2 - a^2(t) ( dr^2 + r^2 d\Omega^2 )
\label{eqn:flrw}
\end{equation}
where $a(t)$ is the scale factor whose dynamical evolution is governed by the Friedmann equations. Here, $d\Omega$ corresponds to angular coordinates, which will be neglected in the remainder of this paper as we will only consider radial motion.

The key point of the cosmological argument is to
rewrite the FLRW metric in what they term ``observer-dependent coordinates'' \cite{2009IJMPD..18.1889M}, initially defining the proper radius as $R(t) \equiv a(t) r$, 
where $r$ is the radial coordinate in the FLRW metric (Equation~\ref{eqn:flrw}),
such that
\begin{equation}
ds^2 = 
\Phi \left[ 1 + \left( \frac{R}{R_h} \right)\Phi^{-1} 
\frac{\dot{R}}{c}\right]^2 c^2 dt^2 - \Phi^{-1} dR^2
\label{eqn:melia}    
\end{equation}
where $R_h$ is the distance to the Hubble sphere \cite[termed the ``gravitational horizon'' by][]{2009IJMPD..18.1889M} given by
\begin{equation}
R_h = \frac{c}{H} = c \frac{a}{\dot{a}} \ .
\label{eqn:hubble}
\end{equation}
 Here, $H$ is the Hubble constant, and 
\begin{equation}
\Phi = 1 - \left( \frac{R}{R_h} \right)^2
\label{eqn:phi}
\end{equation}
At this point, it should be noted that, based upon this representation of the cosmological space-time, \cite{2009IJMPD..18.1889M} conclude that $R_h$ represents a previously unrecognised cosmological horizon which, with other arguments, implies that the universe possess linear expansion, such that $R_h = c t$. 
Whilst it has been claimed that such an expansion history represents a better explanation of observed astronomical phenomenon \cite[e.g.][]{2013AA...553A..76M,2018EL....12359002M,2018MNRAS.481.4855M}
the underlying interpretation has a number of mathematical and physical issues that question its validity \cite[e.g.][]{2010MNRAS.404.1633V,2012MNRAS.423L..26L,2012MNRAS.425.1664B,2013MNRAS.432.2324L,2014MNRAS.442..382M,2016MNRAS.460L.119K,2016MNRAS.460..291L}; I will not revisit this discussion here. 

Returning to \cite{2019IJMPA..3450055M}, the starting point of their argument is to define the 4-momentum of a particle to be
\begin{equation}
p^\mu = \left( \frac{E}{c} , p^R \right)
\label{eqn:momentum}
\end{equation}
where $E$ is the particle energy and $p^R$ is the radial component of the momentum, given by $p^R = m \dot{R}$,
corresponding to a comoving particle with $r=const$ in the FLRW space-time (Equation~\ref{eqn:flrw}); here again I consider only radial motion and neglected the angular components.
By considering the normalisation of the 4-momentum, given by the invariant contraction $p^\mu p_\mu$, such that
\begin{equation}
\Phi \left[ 1 + \left( \frac{R}{R_h} \right)\Phi^{-1} 
\frac{\dot{R}}{c}\right]^2 \left( \frac{E}{c}  \right)^2- \Phi^{-1} (m\dot{R})^2 = K^2
\label{eqn:norm} 
\end{equation}
where $K$ is an undetermined constant. 
Noting that, for a comoving particle and the definition of $R$,
\begin{equation}
R_h = c\frac{a}{\dot{a}} = c \frac{R}{\dot{R}}
\label{eqn:simple}
\end{equation}
then
\begin{equation}
m \dot{R} = m c \left( \frac{R}{R_h} \right)
\label{eqn:ehrel}
\end{equation}
and the energy term in Equation~\ref{eqn:momentum} can be expressed as
\begin{equation}
E^2 = (mc^2)^2 \left[ 1 - \left( \frac{R}{R_h} \right)^2 \right] 
\left( \frac{K}{mc} \right)^2 + (mc^2)^2 \left( \frac{R}{R_h}\right)^2
\label{eqn:fullenergy}
\end{equation}
Clearly, $E( R_h ) = mc^2$ and through an argument with regards to energy conservation along a geodesic, \cite{2019IJMPA..3450055M} deduces that the constant, $K=mc$, and so the energy term is $E=mc^2$ all along the particles trajectory. 

However, given that Equation~\ref{eqn:fullenergy} consists of two parts, \cite{2019IJMPA..3450055M} further interprets this expression a transition from a binding energy when the particle is at the origin, into a purely kinetic energy when the particle reaches the gravitational horizon, with the total energy remaining constant as $E=mc^2$.

\section{No Cosmic Basis for $E=mc^2$}
\label{sec:nobasis}

There are several problems associated with the analysis presented in the previous section. Before considering these, it's important to remember key features of general relativity, namely whilst that the components of any quantity represented by a 4-vector will be dependent upon the coordinate system it is represented in, the magnitude of the 4-vector  will be invariant between coordinate systems. 

Central to following discussion are the properties of the 4-velocity, $u^\mu$, namely
\begin{equation}
u^\mu u_\mu = g_{\mu\nu} u^\mu u^\nu = c^2
\label{eqn:4vel}
\end{equation}
and 4-momentum, defined by $p^\mu = m u^\mu$, such that
\begin{equation}
p^\mu p_\mu = g_{\mu\nu} p^\mu p^\nu = (m c)^2
\label{eqn:4mom}
\end{equation}
where the right-hand side in each case represents the invariant quantities in each case.
Additionally, the energy of a particle is not an absolute quantity, but is relative and dependent upon the observer \cite[see][]{1994AmJPh..62..903N}. More specifically, the value of a particle's energy as determined by an observer is calculated 
 through the projection of the particle's 4-momentum onto the 4-velocity of the observer \cite[see e.g.][]{hartle:2003}, such that
\begin{equation}
E = g_{\mu\nu} p^\mu u^\nu
\label{eqn:energy}
\end{equation}
The key point here is that, in general, the components of the 4-momentum of a particle only has physical meaning when they are projected into an observer's laboratory frame. Hence the association of the first component of the 4-momentum given in Equation~\ref{eqn:momentum} with the energy of a particle is incorrect in general space-times.

The first step in approaching this correctly is to determine the components of the 4-velocity for a comoving particle in the FLRW space-time (Equation~\ref{eqn:flrw}) in the metric given in Equation~\ref{eqn:melia}. This is 
\begin{equation}
u^\mu = \left( u^t , u^R \right) = \left( u^t , \dot{R} \right)
\label{eqn:4vel_new}
\end{equation}
where the component $u^t$ is determined from the 4-velocity normalisation given in Equation~\ref{eqn:4vel}, so
\begin{equation}
\Phi \left[ 1 + \left( \frac{R}{R_h} \right)\Phi^{-1} 
\frac{\dot{R}}{c}\right]^2 (u^t)^2- \Phi^{-1} (\dot{R})^2 = c^2
\label{4vel_new_norm}    
\end{equation}
This can be written as
\begin{equation}
u^t = c \frac{ \sqrt{ \Phi + \left( \frac{\dot{R}}{c} \right)^2 }}
{\left[ \Phi + \left(\frac{R}{R_h}\right)\frac{\dot{R}}{c}     \right]}
\label{eqn:long}
\end{equation}
This messy expression is simplified through the use of Equation~\ref{eqn:simple} for a comoving particle, 
with the result at the 4-velocity is given by
\begin{equation}
u^\mu = \left( c , \dot{R} \right)
\label{eqn:4vel_new2}
\end{equation}
Note that this result is readily apparent from transforming the 4-velocity of a comoving particle in the FLRW coordinates to the space-time given by Equation~\ref{eqn:melia}. 
The corresponding 4-momentum of the particle is given by
\begin{equation}
p^\mu = m u^\mu = m \left( c , \dot{R} \right)
\label{eqn:4mom_super}
\end{equation}
Using Equation~\ref{eqn:fullenergy}, the particle's 4-momentum can be projected onto its own 4-velocity to determine its energy as seen in its own reference frame, finding, as expected, that $E=mc^2$. Hence, the conclusion at this point is that the first term in 4-momentum presented in the RHS of Equation~\ref{eqn:momentum} cannot be simply interpreted as the energy of the particle. 

An examination of Equation~\ref{eqn:fullenergy} reveals that radial component of the 4-momentum, $p^R$, increases with $R$, such that at $R_h$, 
\begin{equation}
p^\mu(R_h) = m ( c , c ) \ 
\label{eqn:4mom_rh}
\end{equation}
and at distances greater than $R_h$, the spatial component in the 4-momentum exceeds the speed of light.
Does this agree with the assertion in \cite{2019IJMPA..3450055M} that the energy of the particle transitions from a rest energy to a completely kinetic quantity at the gravitational radius? As  seen, in projecting the particle's 4-momentum onto its own 4-velocity simply gives $E=mc^2$. 

What if  the particle's 4-momentum is projected onto the 4-velocity of of an observer at the origin? Note that this is not physically legitimate as you can only project vectors onto each other at the same location, but maybe this will reveal the energy transition outlined in \cite{2019IJMPA..3450055M}.
Given that the 4-velocity of an observer at the origin is given by $u^\mu=(c,0)$, the projection of a particle with 4-momentum given by 
$p^\mu = m ( c , \dot{R} )$ will again result in a determination of the energy to be $E=mc^2$ as the spatial component of the 4-momentum does not contribute to the calculated energy. Hence, there is no basis for the claim that there is a transition of the energy of the particle into being purely kinetic when it reaches, passes through, and travels beyond the gravitational horizon.
It is worth noting at this point that these results are applicable for all comoving observers represented as $r=const$ in the FLRW metric (Equation~\ref{eqn:flrw}), irrespective of the expansion history given by the scale factor, $a(t)$.

Finally, \cite{2019IJMPA..3450055M} comments that, in determining the cosmological basis for $E=mc^2$, it is important to consider who the observer is, suggesting that the view point in the FLRW metric (Equation~\ref{eqn:flrw}) and the ``observer dependent'' coordinates (Equation~\ref{eqn:melia}) are distinctly different. However, this is clearly incorrect as, given that these two metrics are representations of the same underlying space-time, the observed quantities of equivalent observers in the two coordinate systems must be identical. 
  
\section{CONCLUSION}
\label{sec:conclusions}
In this paper, I have examined the recent claim that the relativistic equation, $E=mc^2$, has a cosmological basis, representing a binding energy that needs to be overcome for a particle to reach the gravitational horizon of the universe. I have shown that these conclusions arise from an erroneous use of relativistic conceptions for particles moving in expanding space-time, and, with the correct approach, I demonstrated that $E=mc^2$ arises from the place it always has, from the invariant quantity associated with the relativistic 4-momentum.

\begin{acknowledgements}
GFL acknowledges support from a Partnership Collaboration Award between the University of Sydney and the University of Edinburgh, and appreciates the hospitality of the Royal Observatory, Edinburgh where he found the peace and quiet to think about the universe.
\end{acknowledgements}

\bibliographystyle{spphys}       
\bibliography{cosmology}   

%
%

\end{document}